\pdfoutput=1

\documentclass[twocolumn]{aastex62}
\usepackage{CJK}

\newcommand{\ssst}{\scriptscriptstyle}
\newcommand{\E}[1]{\times 10^{#1}}

\newcommand{\s}{\,{\rm s}}      
\newcommand{\ps}{\,{\rm s}^{-1}}

\newcommand{\cm}{\,{\rm cm}}

\newcommand{\kpc}{\,{\rm kpc}} 
\newcommand{\pc}{\,{\rm pc}}
\newcommand{\erg}{\,{\rm erg}}

\newcommand{\nel}{n_{e}}        
\newcommand{\NH}{N_{\rm H}}

\newcommand{\nH}{n_{\ssst\rm H}}

\newcommand{\XMMN}{{\sl XMM-Newton}}

\newcommand{\du}{d_{4}}

\newcommand{\snr}{G7.7$-$3.7}
\newcommand{\Dou}{{\it Nan-Dou}}

\graphicspath{{./}{figures/}}

\submitjournal{ApJL}

\begin{document}

\begin{CJK*}{UTF8}{bsmi}

\title{G7.7-3.7: a young supernova remnant probably associated 
with the guest star~in 386 CE (SN~386)}

\author{Ping Zhou (周平)}
\affil{Anton Pannekoek Institute for Astronomy, University of Amsterdam, Science Park 904, 1098 XH Amsterdam, The Netherlands}
\email{p.zhou@uva.nl}
\affil{School of Astronomy and Space Science, Nanjing University,
163 Xianlin Avenue, Nanjing, 210023, China}

\author{Jacco Vink}
\affil{Anton Pannekoek Institute for Astronomy, University of Amsterdam, Science Park 904, 1098 XH Amsterdam, The Netherlands}
\affil{GRAPPA, University of Amsterdam, Science Park 904, 1098 XH Amsterdam, The Netherlands}
\affil{SRON, Netherlands Institute for Space Research, Sorbonnelaan 2, 3584 CA Utrecht, The Netherlands}

\author{Geng Li (黎耕)}
\affil{National Astronomical Observatories, Chinese Academy of Sciences, 20A Datun Road, Chaoyang District, Beijing 100101, China}
\affil{School of Astronomy and Space Science, University of Chinese Academy of Sciences, No.19A Yuquan Road, Shijingshan District, Beijing 100049, China}

\author{Vladim{\'i}r Dom{\v c}ek}
\affil{Anton Pannekoek Institute for Astronomy, University of Amsterdam, Science Park 904, 1098 XH Amsterdam, The Netherlands}
\affil{GRAPPA, University of Amsterdam, Science Park 904, 1098 XH Amsterdam, The Netherlands}



\begin{abstract}
Although the Galactic supernova rate is about 2 per century, only few supernova remnants are associated with historical records. There are a few ancient Chinese records of ``guest stars" that are probably sightings of supernovae for which the
associated supernova remnant is not established. Here we present an X-ray study of the 
supernova remnant G7.7$-3.7$, as observed by \XMMN, and discuss its probable association with the guest star of 386~CE.
This guest star occurred in the ancient Chinese asterism \Dou, which is part of
{\it Sagittarius}.
The X-ray morphology of  G7.7$-3.7$ shows an arc-like feature in the SNR south, which is
characterized by an under-ionized plasma with sub-solar abundances, 
a temperature of $0.4$--0.8~keV, and a density of  $\sim  0.5(d/4~\kpc)^{-0.5}\cm^{-3}$.
A small shock age of $1.2\pm 0.6 (d/4~\kpc)^{0.5}$~kyr is
inferred from the low ionization timescale of $2.4^{+1.1}_{-1.3}\E{10}~\cm^{-3}\s$ of the X-ray arc.
The low foreground absorption ($\NH=3.5\pm0.5\E{21}~\cm^{-2}$)
of \snr\ made the supernova explosion visible to the naked eyes
on the Earth. 
The position of  \snr\ is consistent with the event of 386~CE, and the X-ray properties suggest that also
its age is consistent. 
Interestingly, the association between \snr\ and guest star~386 
would suggest the supernova to be a low-luminosity supernova, 
in order to explain the not very long visibility (2--4 months) of 
the guest star.

\end{abstract}



\section{Introduction} \label{sec:intro}

The energetic explosions of supernovae (SNe) are among the brightest
events in a galaxy, but have a short-lived brightness which quickly 
declines a few magnitudes within a year.
A SN rate of about 2 per century \citep{reed05,diehl06} indicates tens of 
Galactic SNe were exploded in the past two millenniums.
However, only a few nearby SNe were observed 
with naked eyes and recorded by ancient astronomers.
The debris of these historical transient events are now among
the youngest members of Galactic supernova 
remnant (SNR) population.
Historical SNRs play an important role in mutual 
understanding of SNR physics and ancient Astronomy, and have received
most ample attention.
The well recorded group of historical SNRs includes Type Ia SNRs 
such as Tycho's (SN~1572) and Kepler's SNRs (SN~1064), SN~1006,  
as well as Type II SNRs Crab (SN~1054, a.k.a. guest star 
``Tian-Guan'') and 3C58 \citep[likely SN~1181, see][and references therein]{green03}.
Among them, 3C58 is regarded as a probable remnant of SN~1181, 
as the slow expansion of the pulsar wind nebula and the coolness
of the pulsar are better explained if the SNR is older than
2000 yr \citep[e.g.,][]{bietenholz01,slane02,fesen08}.

The long duration transients that occurred more than a thousand years ago,
mainly in Chinese records \citep{xi55,hsi57,xi65}, may 
provide crucial clues for finding other young SNRs.
It is more difficult to establish a clear association 
between SNRs and the records, given the uncertainties of derived 
SNR ages and the vague position in ancient texts. 
Despite the difficulties, RCW~86 has been considered as a probable 
SNR of  the guest star observed in 185 CE that was visible for 
20 or 8 months \citep[e.g.,][]{westerlund69,clark77, vink06a}.
RX J1713.7-3946 is suggested to be the SNR of  the guest star  
in 393~CE \citep[e.g.,][]{wang97}.
Guest star 393 appeared during {\it Jin} (晉) Dynasty, 
when two other long-duration guest stars were reported 
in 386~CE (2--4 months) and 369~CE \citep[7--9 months, see][and references therein]{xi55, xi65}. 

The earliest available record of guest star 386 was found in 
\underline{Song Shu} (\underline{宋書}) : 
{\it 3rd month, 11th year of the Tai-Yuan 
reign-period of Jin dynasty, there was a guest star
at 
\Dou\ (南斗) until
6th month when it was extinguished} \citep[see][]{hsi57, clark77}.
The records are very much similar in the later text sources
\underline{Jin Shu} (\underline{晉書}), \underline{Tong Zhi} (\underline{通志}), 
\underline{Wen Xian Tong Kao} (\underline{文獻通考}), but
shortened, probably because they transcribed the same earlier record.
The ancient Chinese asterism \Dou\ (a.k.a. {\it Dou}, Southern
Dipper or Dipper), as part of Sagittarius, is composed of six stars in 
the Galactic coordinates range
$l=6\fdg{9}$--$10\fdg{0}$, $b=-1\fdg{6}$--$-15\fdg{4}$.
\Dou\ or {\it Dou} can also refer to a lunar mansion
that covers a range of right ascension and a few asterisms including
\Dou\ asterism itself \citep[R.A=$17^{\rm h} 15^{\rm m}$ to 
$18^{\rm h} 50^{\rm m}$; see][]{stephenson02}. 
It is an enigma whether \Dou\ meant an asterism or
a lunar mansion for guest star 386, but 
if a young remnant with a consistent age is detected in \Dou\ 
asterism, it would  provide a support for the former meaning.

SNR G11.2$-$0.3 was initially regarded as the prime candidate remnant
of the guest star \citep{clark77, green03}, but was recently ruled out,
since it cannot have been visible by the naked eye, given its 
distance and large extinction \citep[$A_V=16\pm2$,][]{borkowski16}.
Another candidate SNR, \snr, is located in \Dou\ asterism
\citep{stephenson02}.
However, the age of it is unclear limited by the
very few studies and observations.
It has a shell-like morphology in the radio band 
\citep{milne86,milne87, dubner96}.
The empirical $\Sigma$--$D$ (radio surface-brightness -- distance) relation, which has considerable scatter,
provides a very rough SNR-distance of 
$\sim 3.2$--6~kpc \citep{milne86,pavlovic14}.
The distance was suggested to be closer than 6~kpc, otherwise,
the SNR would be unusually far out of the Galactic plane \citep{milne86}.

An X-ray imaging study with short-exposure \XMMN\ data 
in 2005  (partially covering the SNR) showed that 
\snr\ contains an X-ray feature in the south  and 
a point-like source in the north-east \citep{giacani10}.

Motived by the need to link  young SNRs with early 
historical records, and the 
uncertain association of the SNR 
with the guest star
of 386, 
we performed an analysis of the SNR~\snr.
We report here that \snr\ is indeed a young SNR in \Dou, and suggest
that it is the remnant of the guest star 386 (SN~386), according
our X-ray imaging and spectroscopic analysis of this remnant.

\section{Data}
\snr\ was observed with \XMMN\ in 2005 
(obs. ID:0304220401, PI: E.\ Gotthelf)
and 2012
(obs. ID:0671170101, PI: M.\ Smith). 
The whole SNR was covered by the pn and MOS cameras 
of the 2012 observation, which were operating in full 
frame mode with a thin filter.
The 2005 observation was targeted to the nearby point source
AX~J1817.6$-$2401, and pn/MOS cameras were in full 
frame/small window mode with a medium filter.
For this observation, the X-ray emitting region of \snr\ was fully 
covered by MOS2 camera at an off-axis region, partially
covered by MOS1 camera, but were missed by the pn camera.
Therefore, we use  the MOS2 data in 2005
and  all the MOS and pn data in 2012.
The total exposure 
times
were 13 and 40 ks, respectively,
but both observations suffered soft proton flares
in most of the observation time.
After removing the high background periods from the
events, the screened exposure time in 2012 is 
6.3, 7.5 and 6.6 ks for pn, MOS1 and MOS2, respectively,
and the net exposure time of the MOS2 data in 2005 is only 
2.9~ks.

The \XMMN\ data were reduced using the Science Analysis 
System software (SAS, vers.\ 16.1.0)\footnote{https://www.cosmos.esa.int/web/xmm-newton/download-and-install-sas}.
We used
XSPEC (vers.\ 12.10.0c)\footnote{https://heasarc.gsfc.nasa.gov/xanadu/xspec/} 
with AtomDB 3.0.9\footnote{http://www.atomdb.org/} and and SPEX (vers. 3.0.4) with SPEXACT 3.04 atomic tables \citep{Kaastra96} for a comparative spectral analysis.
The fitting statistics method employed was C-statistics \citep{Cash79}.

We also retrieved the 1.4 GHz radio continuum data 
from NRAO VLA Sky Survey \citep[NVSS,][]{condon98}
for a comparison with the X-ray data.

\section{X-ray analysis of \snr} 

\begin{figure}
\epsscale{1.1}
\plotone{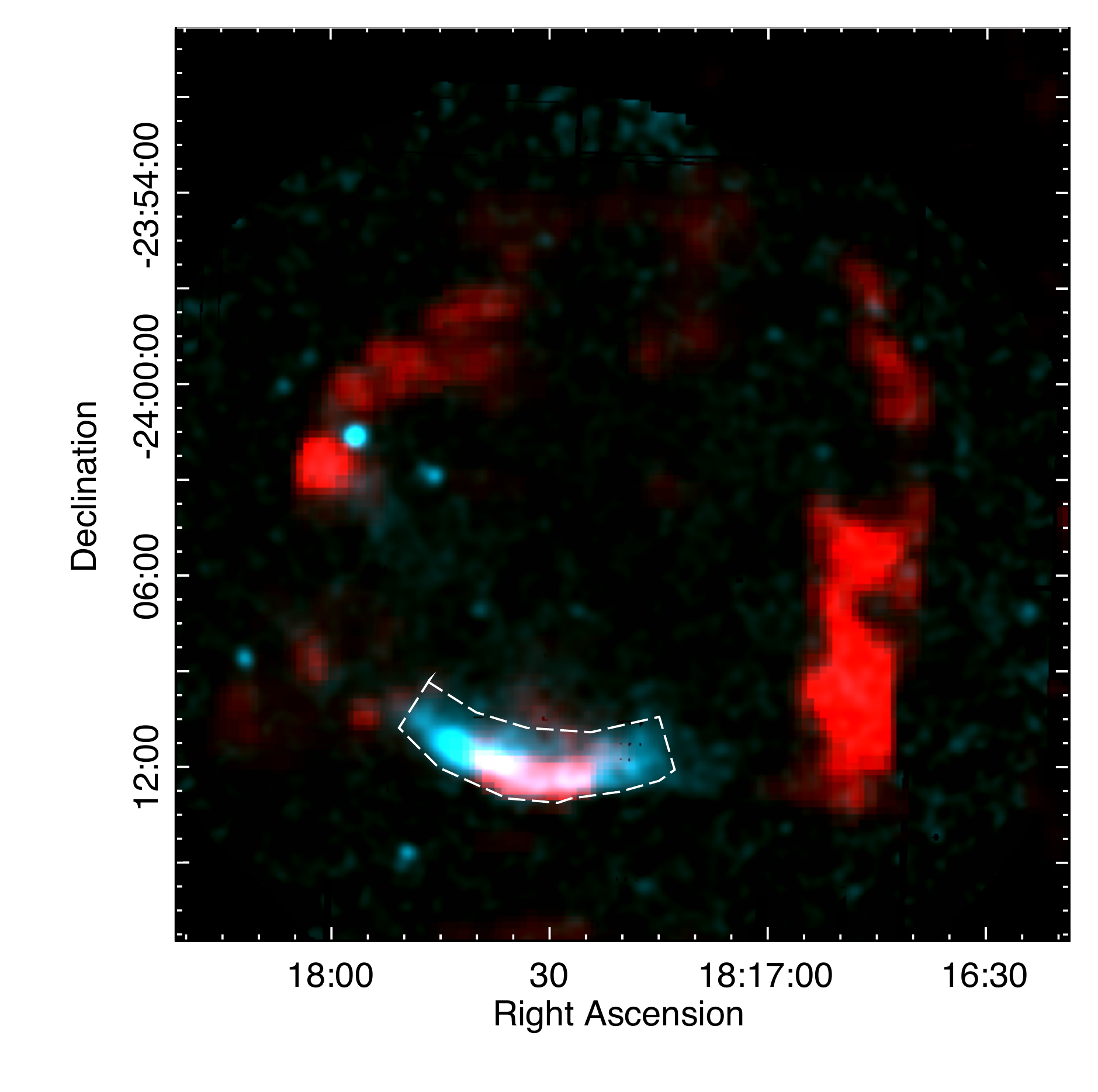}
\caption{
A composite image of \snr\ in X-ray (cyan) and
1.4~GHz radio bands (red). 
The X-ray image was made using the \XMMN\
pn and MOS data in 2012 (obs. ID: 0671170101).
It was vignetting-corrected, instrumental-background-subtracted
and adaptively smoothed to reach a S/N=10.
\label{fig:img}}
\end{figure}

\begin{figure*}
\epsscale{1.1}
\plottwo{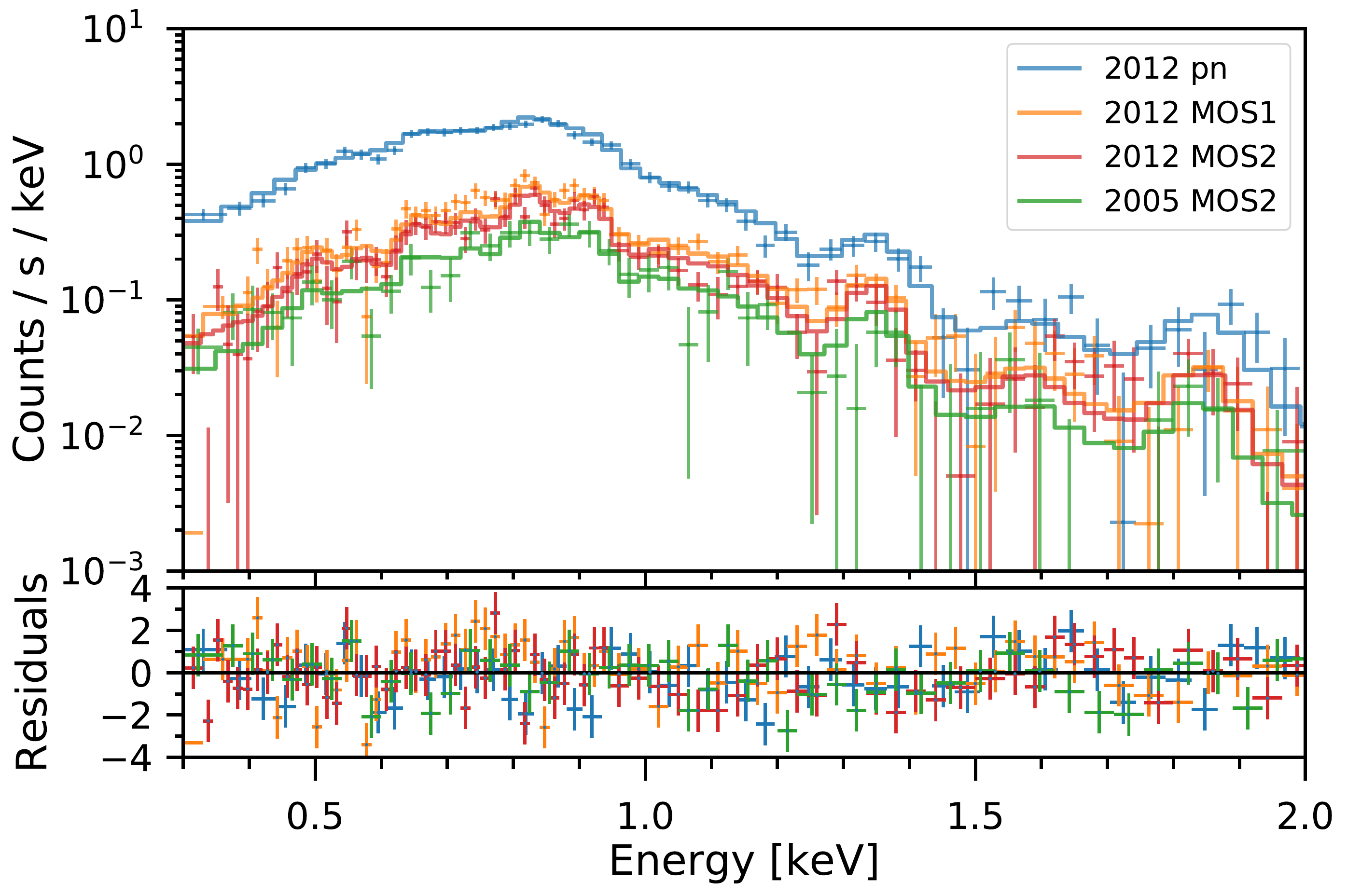}{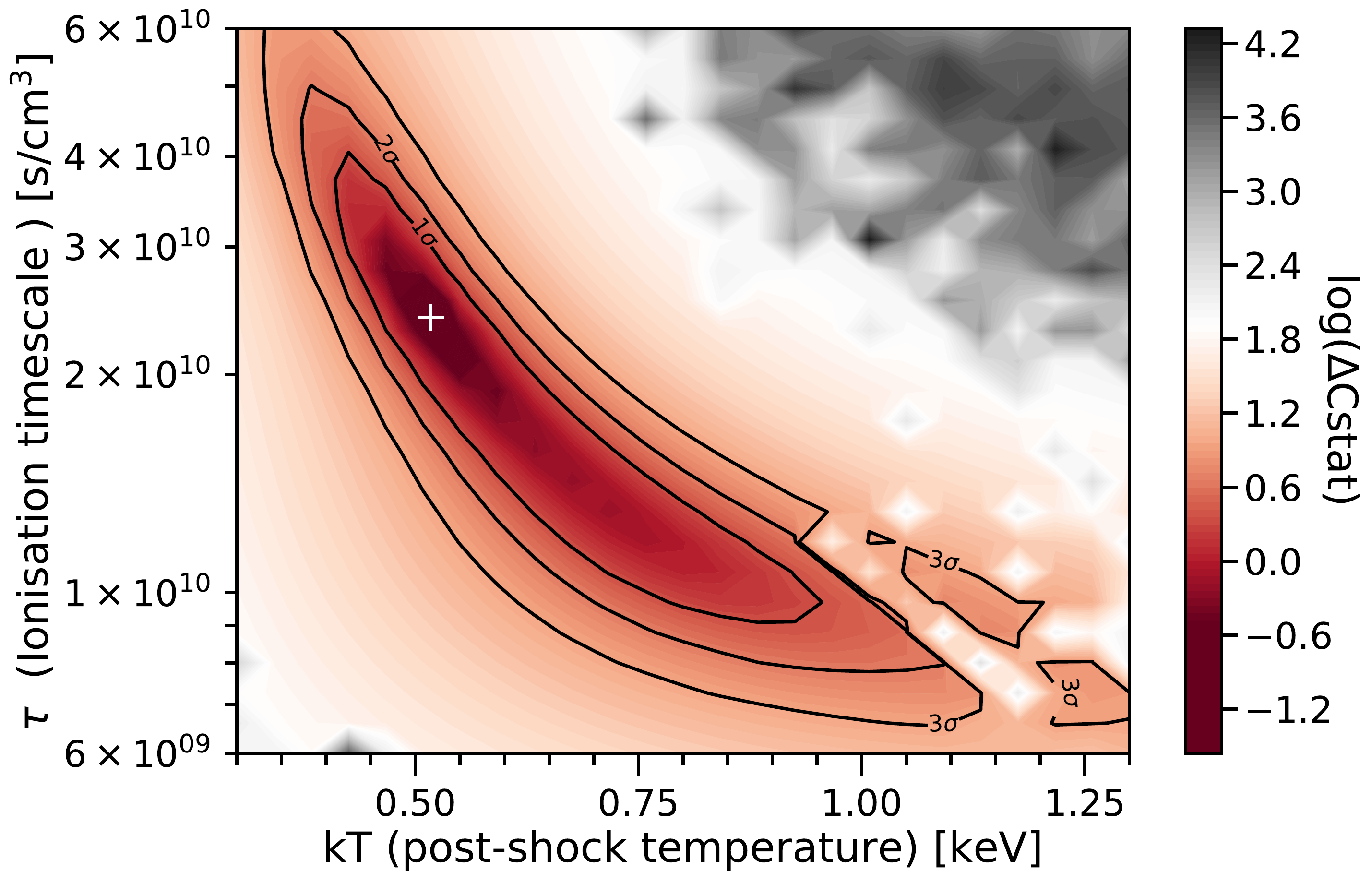}
\caption{Left: The spectra of the southern arc
fitted with an absorbed {\it neij} model.
The residuals = (data-model)/model.
Right: The correlation between the best-fit ionization timescale $\tau$
and post-shock temperature $kT$, overlaid with the 1, 2, and 3-$\sigma$ 
confidence levels (SPEX results).
\label{fig:spec}}
\end{figure*}

The X-ray image of \snr\ in 0.3--3 keV reveals an
arc-like feature in the south (see Figure~\ref{fig:img}),
which is unlike the radio emission that delineates the SNR's 
boundary \citep[see][for a study in radio band]{dubner96}.
There seems to be an anti-correlation of the brightness
between X-ray and radio emission along the arc:
the radio emission is brightest in the arc center, where
the X-ray emission is fainter.
Although the southern arc is the only bright X-ray feature
in \snr, very dim, diffuse X-ray emission is detected
inside the eastern radio boundary.

To study the properties of the X-ray arc of \snr,
we extracted spectra from the region surrounding the
arc (see the dashed region in Figure~\ref{fig:img}).
The background is subtracted from a nearby source-free region.
Four groups of spectra are extracted, from the pn/MOS1/MOS2
observation in 2012, and MOS2 observation in 2005.
The left panel of figure~\ref{fig:spec} shows the spectra optimally 
binned \citep{kaastra16} using the {\it Obin} command in SPEX. 
The detected X-ray photons mainly have an energy below 2 keV.
A He-like Mg line is clearly detected at 1.33~keV
and a He-like Si bump is shown at $\sim 1.85$ keV,
indicating a thermal origin of the emission.
We also tested an absorbed power-law model plus 
a Gaussian line at 1.33~keV,
which results in unacceptable fit (reduced chi-squared $\chi^2_\nu=2.4$ in XSPEC)
and an extremely steep photon index ($9.5$).
This further supports that the X-ray spectra are dominated by
thermal emission.
We therefore applied an absorbed non-equilibrium 
ionization (NEI) model ({\it vnei} in XSPEC and {\it neij} in SPEX) to 
jointly  fit the 4 spectra above 0.3~keV. 
The NEI model was employed with the following free parameters: 
electron temperature $kT$, ionization timescale 
$ \tau = \nel t $, 
abundances of elements O, Ne, Mg, Si, Fe,
and normalization scaled with the volume  emission measure.
For the analysis with SPEX, we applied the absorption model {\it hot} and solar abundances from \citet{lodders09}.
For XSPEC, we used 
the absorption model {\it tbabs} \citep{wilms00} and the solar abundances 
from \citet{asplund09}.
Each individual spectrum used in XSPEC is adaptively binned to achieve a 
background-subtracted signal-to-noise ratio (S/N) of three.

The absorbed NEI models well describe the spectra, as shown in the left panel of Figure~\ref{fig:spec} (SPEX model).
In spite of differences in the two spectral fitting packages and
binning methods,
we found the best-fit results are similar to each other, except that the XSPEC analysis 
provides  tighter constraints on the parameters, as summarized in Table~\ref{tab:spec}.
Hereafter, we conservatively use the results obtained with SPEX, 
considering that it provides uncertainty ranges mostly covering
those from XSPEC.
The low foreground absorption 
$\NH=3.5\pm 0.5\E{21} \cm^{-2}$
implies that \snr\ is not much absorbed compared 
to many other SNRs toward the Galactic center.
The spectra are characterized by an under-ionized thermal 
component with an electron temperature $kT=0.5_{-0.1}^{+0.3}$~keV 
and sub-solar abundances.
The low abundances suggest that the X-ray emission is from 
shocked interstellar medium rather than from the ejecta.
The electron temperature $kT$ and ionization timescale $\tau$ show a 
degeneracy in spectral fit as shown in the right panel
of Figure~\ref{fig:spec},
with higher fitted $kT$ values accompanied with lower fitted $\tau$.

\begin{center}
\begin{deluxetable}{p{4.cm}cc}
\tabletypesize{\footnotesize}
\tablecaption{X-ray spectral fit results with 1-$\sigma$ uncertainties
\label{tab:spec}
}
\tablewidth{0pt}
\tablehead{
\colhead{Parameter} & XSPEC results & SPEX results
}
\startdata
$N_{\rm H}$ ($10^{21}~\cm^{-2}$)& $3.4\pm 0.4$ &  $3.5\pm0.5$\\
$kT$ (keV) & $0.61^{+0.10}_{-0.09}$ & $0.52^{+0.30}_{-0.08}$\\
O &  $0.30\pm 0.05$ & $0.32_{-0.04}^{+0.09}$\\
Ne & $0.64^{+0.12}_{-0.11}$ & $0.45_{-0.07}^{+0.15}$\\
Mg & $0.43^{+0.11}_{-0.09}$ & $0.48_{-0.09}^{+0.13}$\\
Si & $0.48^{+0.19}_{-0.17}$ & $0.61_{-0.12}^{+0.14}$\\
Fe & $0.56^{+0.12}_{-0.10}$ & $0.70_{-0.12}^{+0.14}$\\
$\tau$ ($10^{10}~\cm^{-3}\s$) & $2.7^{+0.9}_{-0.6}$ & $2.4_{-1.3}^{+1.1}$\\
norm ($\cm^{-5}$) & $2.4^{+1.3}_{-0.8}\times 10^{-3}$ & $3.2_{-1.8}^{+2.2}\E{-3}$\\
C-Stat.\ & 339.66 & 391.69 \\
$\chi^2$/d.o.f    & 345.00/306 &  306.90/239\\
\enddata
\end{deluxetable}
\end{center}

The density of the gas, $\nH$, is calculated using the best-fit 
X-ray volume emission measure (from the parameter 
$norm =10^{-14}/(4\pi d^2) \int \nel\nH dV$)\footnote{https://heasarc.gsfc.nasa.gov/xanadu/xspec/manual/node190.html}, where
$d$ is the SNR distance, $\nel$ and $\nH$ are the electron
and H densities in a volume $V$, and $\nel=1.2\nH$ for fully ionized plasma.
As the spectra are selected from an arc-like region
with a thickness of $2'$, we
assume that the geometry of the X-ray-emitting region is a
thin cap  with apex angle of $52\fdg{8}$ (length is $8'$) 
and a radius the same as the SNR's radius $R=9'$.
We adopt one twelfth $R$ as the thickness of the 
cap, which is the expected thickness of the shell for
a uniform density $\rho_0$ and a shock compression ratio of four
($4\pi R^2\Delta R (4\rho_0)=4\pi/3 R^3 \rho_0$).
The thickness of $1/12R$ is found to be a good approximation, since
the projection effect of the cap can result in a visual 
thickness of $1\farcm6$. The larger point-spread-function 
at an off-axis region of \XMMN\ and our adaptive smoothing 
procedure can further smooth the arc to reach the thickness
of $\sim 2'$.
As a result, the plasma density is estimated to be 
$\nH=0.53^{+0.18}_{-0.15} \du^{-0.5}~\cm^{-3}$,
where $\du = d/(4 \kpc)$ is the distance scaled to 4 kpc.

An SNR age can be inferred from the ionization timescale
$\tau$ if its plasma has not reached ionization equilibrium. 
For under-ionized plasma, $\tau$ is defined as a product of
the electron 
density and the time elapse since the gas is shock 
heated ($\tau=\nel t$).
The shock age $t$ of \snr\ estimated to be
$1.2\pm0.6~\du^{0.5}$~kyr,  given the low ionization timescale 
$\tau=2.4^{+1.1}_{-1.3} \E{10}~\cm^{-3}\s$ and the best-fit H density.
The small shock age implies that \snr\ is a young SNR that 
could possibly have been observed by ancient astronomers.

\snr\ does not show evidence for synchrotron X-ray emission or
bright thermal emission with enhanced metal abundances
in the hard X-ray band,
which is unlike other young/historical SNRs. 
Nevertheless, if this hard component is very faint, it 
could not be easily detected with the $<10~$~ks observations.
Note that the SNR is likely expanding into a low-density 
($\lesssim 0.1~\cm^{-3}$) environment, which explains its 
low X-ray brightness.
The ambient density is likely non-uniform, as indicated 
by the not so spherical radio morphology and the enhanced X-ray emission in the south.

\section{Is G7.7$-3.7$ the remnant from guest star 386?}

\begin{figure}
\epsscale{1.25}
\plotone{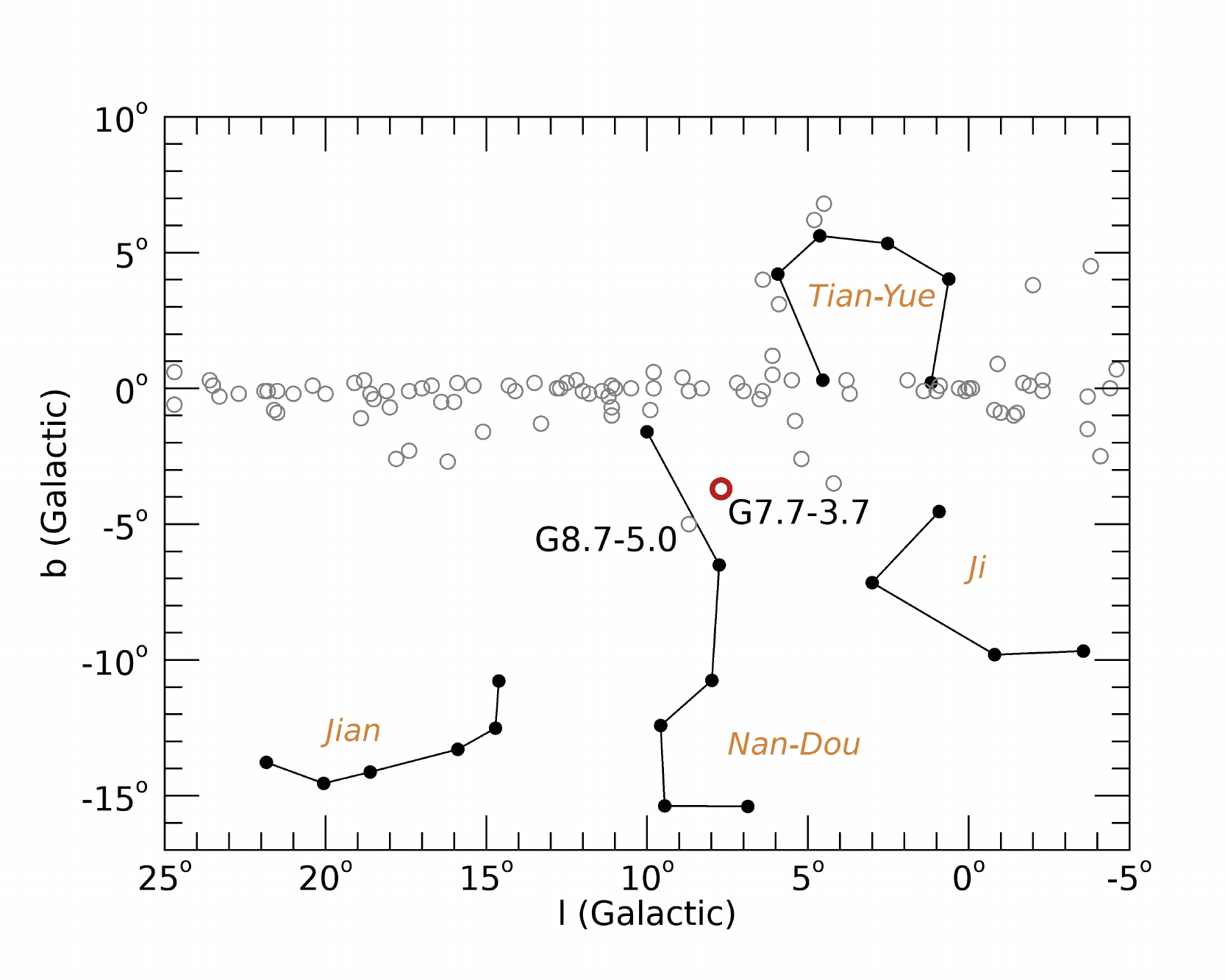}
\caption{
The ancient Chinese asterism \Dou\ and three nearby
asterisms. The known SNRs are shown with open circles. 
\snr\ is denoted with a thick open circle.
\label{fig:starmap}}
\end{figure}

\snr\ is located near the stars \Dou\ {\it V ($\lambda$ Sgr)}
and \Dou\ {\it VI ($\mu$ Sgr)} 
in the ancient Chinese asterism \Dou,
Figure~\ref{fig:starmap} shows \Dou\
and other three nearby asterisms, where the 
known SNRs \citep{green14,ferrand12} are over-plotted for comparison.
\snr\ and G8.7$-$5.0 are the known SNRs nearest to
the center of \Dou.
There are very limited studies of the latter SNR.
It has a diameter of $26'$ in radio band
\citep{reich88,reich90}, slightly larger than
\snr.
We didn't find any information for the age
of G8.7$-$5.0. Moreover, checking the X-ray archival data by
ROSAT, Swift or  \XMMN, we did not find evidence for X-ray emission
from this object. This suggests that  G8.7$-$5.0 is an old remnant,
and, therefore, unlikely to be the remnant of 386~CE.
The youth ($< 2000$~yr) of \snr\ indicates
that it is the more probable remnant of the 
guest star~386 that appeared at \Dou.

The record of guest star 386 is the only historical record 
matching \snr, given the SNR's age (0.6--$1.8\du^{0.5}$~kyr 
for 1-$\sigma$ uncertainty), the position, and expected SN brightness.
Guest star~386 appeared more than  2~months in \Dou, while
other guest stars in the same asterism had short-term
visibility.
The long time the guest star was visible is more probable to \snr, since its SN 
explosion should be bright enough to be observed with naked eyes 
on the Earth (see Section~\ref{sec:visibility}).

It is unlikely that \snr\ is associated with another guest star in that part of the sky:
A guest star was recorded to appear near {\it Kui} of \Dou\ (the four 
southern stars in Figure~\ref{fig:starmap}) on Feb 8, 1011 \citep[\underline{Song Shi}; \underline{宋史}; \underline{Wen Xian Tong Kao};][]{xi55}.
The position does not agree with that of \snr\ and the time is 
too short to be considered as a SN. 
For the duration problem, we also rule out another short-term 
transient event called {\it Xing-Bo} (星孛; brushy star) that appeared in \Dou\ in November, 
1375 \citep[\underline{Guang Dong Tong Zhi}; \underline{廣東通志};][]{xi65}.
Moreover, there is no other matchable record for \snr\ from the 
SN/novae-like events recorded  during 532~BC and 386~CE \citep{hsi57,clark77}.
Therefore, we suggest that guest star 386 is an SN that resulted 
in \snr, given the consistent age and position.
Vice versa, if we except that \snr\ is less than 2000~yr old, there is no other historical record
of the SN event than SN~386.

\section{On the relatively short visibility of the guest star} \label{sec:visibility}

Guest star 386 was only visible with naked eyes during 2--4 months
\citep[Apr 15/May 14--Jul 13/Aug 10, 386,][]{xi65},
shorter than the duration of all other well-studied historical SNe 
with visibilities from half a year to 3 years.
Since \Dou\ was visible in the night sky  from February to 
September 386 at Jiankang (currently Nanjing; the capital of {\it Jin}
at that time), the 2--4 months duration of the guest star is likely intrinsic to the guest star,
and not due to the visibility of that part of the sky before April or after August. The asterism was
visible from January till October, but in January only near dawn and October early in the evening.
Since once a guest star is detected one can keep following it, the disappearance in August means the guest star
was too faint to see with the naked eye.
However,  we cannot completely rule out that the guest star may have escaped attention when the asterism became
first visible during the night in January or February and was only noticed in April.
However, for the present discussion we adopt the more likely possibility that the guest star was only visible for
2--4 months.

The not very long duration suggests that the event was fainter than other
known historical SNe, due to either large visual 
extinction, large distance, or low intrinsic luminosity.
The two former possibilities can be ruled out.
\snr\ is at high Galactic latitude where visual 
extinction $A_V$ is only $1.2\pm 0.2$, as obtained from the 
fitted $\NH$ and the $\NH$--$A_V$ relation \citep[$\NH/A_V=2.87\E{21} 
\cm^{-2}$,][]{foight16}.
If we assume \snr\ was a normal SNe with a peak absolute 
magnitude $M_V=-17$ -- $-19$ at a distance of 3--6~kpc,
the peak apparent magnitude would be  $m_V=M_V+5\log(d/10~\pc)+A_V=-1.7$ -- $-5.6$ mag.
This $m_V$ value based on the assumption of a normal SN would put 
the event among the brightest objects on the night sky at that time 
and visible much longer than 4 months, which conflicts with the 
record of the guest star.

\begin{figure}
\epsscale{1.2}
\plotone{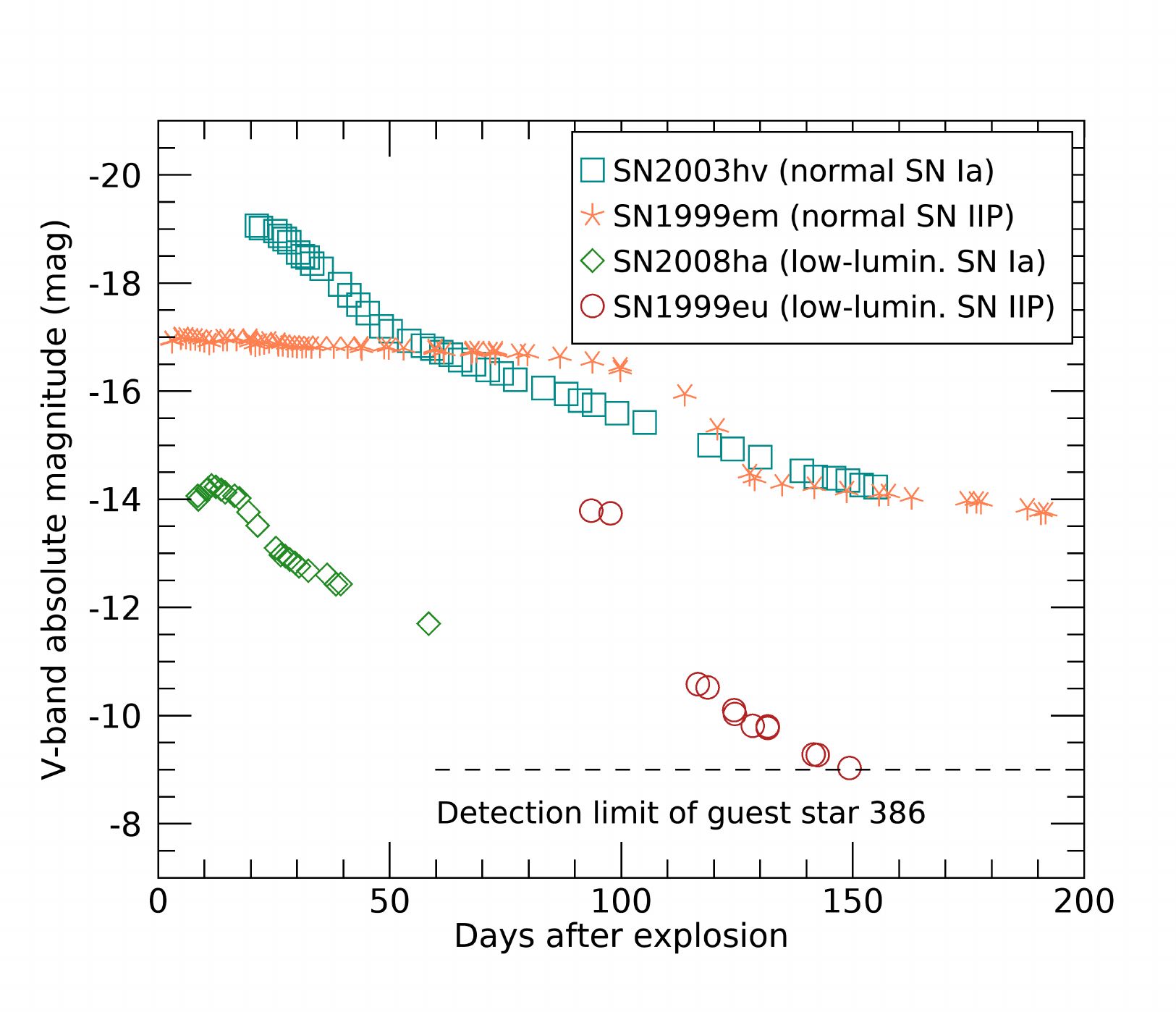}
\caption{
The V-band light curve of a few exemplified SNe: SN 1999eu
\citep{pastorello04}, SN 1999em \citep{hamuy01,leonard03}, SN 2008ha
\citep{foley09}, and SN 2003hv \citep[see][and references therein]{ganeshalingam10, lennarz12}.
The dashed line shows the limiting magnitude of guest star 386 after 2--4 months'  observation, assuming the visual limiting magnitude for
naked eyes was 5 mag.
\label{fig:lc}}
\end{figure}

The absolute magnitude $M_V$ of the guest star shortly before it disappeared 
should be around  $-9$ -- $-8$ mag, since a limiting magnitude for naked 
eyes is 5--6 mag.
The brightness is exceptionally low compared to normal SNe, but may 
be consistent with some low-luminosity SNe (see Figure~\ref{fig:lc}).
A class of low-luminosity Type IIP SNe has a peak $M_V$ 
in the range $-14$--$-16$~mag, and are very faint after 
3--5 month's optical plateau \citep[e.g., as faint as $\sim -9$ mag 
for SN~1999eu,][]{pastorello04}.
Another small class of very low-luminosity SNe belongs to Type Ia.
An extremely sub-luminous case is Type Ia SN~2009ha, which has 
a peak $M_V$ as faint as $-14$~mag, very low photospheric 
velocity and small explosion energy \citep[$2\E{48}\erg$,][]{foley09}.
The relatively high Galactic latitude of
\snr\ might prefer a lower mass progenitor.
However, the explosion kinetic energy of \snr\ is estimated to be 
$E\sim 2.0\E{51}\chi(d/4~\kpc)^{4.5}\erg\ps$,
assuming that the SNR is in Sedov phase, has an age of 1632~yr, where $\chi$ is the deviation of the mean
ambient density from the the arc's preshock density ($\chi <1$ means that
the arc is at a density-enhanced region).
This rules out the possibility of a weak explosion, low-luminosity Type Ia SN.
The distance of \snr\ is likely close to the lower limit of the 
3--6~kpc range 
(with a smaller $E$ of $\sim 5.5\E{50}\chi\erg\ps$ at 3~kpc)
or/and the mean ambient density is much lower than near the arc,
since the low-luminosity Type IIP SNe are suggested to have a relatively
low explosion energy \citep[1--$9\E{50}~\erg$; see Table 5 in][and references therein]{lisakov18}.
Beside the explanation of a very low-luminosity Type IIP SN, there is
a possibility that the extinction of the explosion site is
much higher than at the X-ray arc, which could be tested with 
future observations.

Our last remark is about the intriguing properties of \snr.
If the SNR is indeed the debris of the guest star 386 as we suggested,
it would be the faintest historical SNR in the X-ray band,
with an X-ray flux of $2.3\E{-12} \erg~\cm^{-2}\ps$
in 0.3--2 keV. 
Moreover, \snr\ might be the only known historical SNR resulted from a very low-luminosity
SN, and, therefore, of particular value for studying the evolution and 
origin of this small group of SNe \citep[$<4$--5\% of all Type II SNe,][]{pastorello04}.

\section{Conclusion}

We report that \snr\ is a $1.2\pm 0.6 \du^{0.5}$~kyr young SNR 
probably associated with the guest star of 386~CE, given the consistent age and position, and given that there are no
other compatible historical records that could be associated with this SNR.
The SNR has an non-uniform X-ray distribution, with the X-ray emission only bright 
in the southern arc.
The spectroscopic analysis shows that the shocked plasma is under-ionized
($2.4^{+1.1}_{-1.3}\E{10}~\cm^{-3}\s$) with subsolar abundances.
The association between \snr\ and guest star~386  would suggest 
the supernova to be a low-luminosity supernova, possibly a low-luminosity
Type IIP, in order to explain the not very long visibility 
(2--4 months) of  the guest star.
So far only limited observations and studies exist for \snr.
Given the peculiar properties of the SNR, and its probable association with the guest star of 386~CE we hope that this situation will
improve in the near future.

\begin{acknowledgements}
We are grateful to Shuai-Jun Zhao for a research on available records
of guest star 386.
P.Z. acknowledges the support from the NWO Veni Fellowship, grant no.\ 639.041.647 and NSFC grants 11503008 and 11590781.
G.L. thanks the support from the Youth Innovation Promotion
Association, CAS (grant no. 2016053).
\end{acknowledgements}

\software{
XSPEC \citep{arnaud96},
SPEX (Kaastra et al.\ 1996),
SAS,
DS9\footnote{http://ds9.si.edu/site/Home.html},
Stellarium (vers.\ 0.18.0)
}

\end{CJK*}
\end{document}